\documentclass[
  reprint,
  aps,
  prl,
  superscriptaddress,
  longbibliography
]{revtex4-2}
\usepackage{seqsplit}

\usepackage{graphicx}
\usepackage{subcaption}
\usepackage{placeins}
\usepackage{amsmath,amssymb}
\usepackage{bm}
\usepackage{booktabs}
\usepackage{siunitx}
\usepackage[version=4]{mhchem}
\usepackage[colorlinks=true,allcolors=black]{hyperref}
\usepackage{caption}
\usepackage[final]{microtype}
\usepackage{ragged2e}
\makeatletter
\long\def\@makecaption#1#2{%
  \vskip\abovecaptionskip
  \parbox{\linewidth}{\small\justifying #1.~#2\par}%
  \vskip\belowcaptionskip}
\makeatother

\begin{document}


\title{Resolving Spin-Phonon Relaxation Pathways \\ in Molecular Qubits via Regularized Regression}

\author{Sayan Banerjee}
\email{sbanerjee@utk.edu}
\affiliation{Department of Chemistry,
             University of Tennessee--Knoxville,
             Knoxville, TN 37923, USA}

\begin{abstract}
Designing molecular qubits requires controlling the spin–lattice relaxation pathways that fundamentally limit the coherence time. For spin-1/2 molecular qubits, significant discrepancies remain between theoretical predictions and experimental measurements of spin-phonon relaxation pathways, with theory often overestimating the role of low-frequency vibrational modes. Here, we present an alternative first-principles approach augmented with regularized regression that identifies spin-phonon relaxation pathways in an automated fashion without ad hoc mode selection. Applied to Cu porphyrins spin-1/2 molecular qubits, the method successfully reproduces experimental trends in relaxation times and predicts the curvature in the relaxation-vs-temperature profile. The $g$-tensor time series reveal distinct system-specific autocorrelation functions and spectral-density profiles despite the qubits' structural similarity. Further, regression between $g$-tensor time series and mode-projected vibrational-amplitude time series obtained from molecular dynamics yields linear ($\alpha_k$) and bilinear ($\beta_{kl}$) mode-coupling contributions to each $g$-tensor component. Because the method is based on atomic displacements sampled directly from molecular dynamics, it naturally incorporates anharmonic effects and does not require the harmonic approximation. This framework puts forward a regression-driven, mode-resolved, and coupling-order-separated spin-phonon analysis as a general strategy for predicting and engineering longer-lived molecular qubits.

\end{abstract}

\maketitle

\section{Introduction}
Molecular transition-metal complexes provide a chemically programmable platform for quantum information science, where coordination geometry, ligand field, and vibrational structure can be tuned to engineer spin dynamics.~\cite{wasielewski2020exploiting,gaitaarino2019molecular,qiu2023enhancing} Two metrics are central to this effort: long coherence times ($T_2$) and slow spin--lattice relaxation ($T_1$). Because $T_2$ is fundamentally bounded by $T_1$, extending $T_1$ is a prerequisite for coherence engineering.~\cite{atzori2019second,mirzoyan2020dynamic,lunghi2023spinphonon} In molecular qubits, $T_1$ is governed by spin--phonon coupling, making it essential to identify which vibrational motions couple most strongly to the electron spin and how those couplings drive relaxation.~\cite{mirzoyan2020dynamic,kazmierczak2021impact,lunghi2020limit,lunghi2022toward,lunghi2019phonons,EscaleraMoreno2017} The relevant mechanisms include direct one-phonon processes and Raman two-phonon processes; resolving these pathways at the atomistic level is therefore a key requirement for rational molecular-qubit design.~\cite{lunghi2023spinphonon}


A particularly important testbed is the family of spin-$\tfrac{1}{2}$ Cu(II) porphyrin qubits, for which extensive experimental characterization is available \cite{mirzoyan2020dynamic,kazmierczak2021impact,kazmierczak2024chemsci,lohaus2026jacs,Yu2019,vonKugelgen2021,Mullin2024}. Across this series, discrepancies have persisted between theoretically predicted and experimentally inferred relaxation pathways, with theory often overestimating the importance of low-frequency ruffling modes relative to bond-stretching modes.~\cite{mirzoyan2020dynamic,kazmierczak2024chemsci} Resonance Raman measurements across the closely related \ce{Cu(II)} porphyrin qubits CuOEP, CuTPP, and CuTiPP instead point to bond-stretching modes as the dominant spin--phonon channel,~\cite{kazmierczak2024chemsci} while complementary inelastic-neutron and EPR measurements on CuOEP reveal a temperature-dependent crossover between low-frequency lattice modes and higher-frequency optical phonons.~\cite{lohaus2026jacs} This mismatch highlights a broader challenge in current theory: identifying relaxation-active modes in systems where anharmonicity and multi-mode couplings are expected to play a significant role.

\begin{figure*}[t]
\centering
\includegraphics[width=1\textwidth]{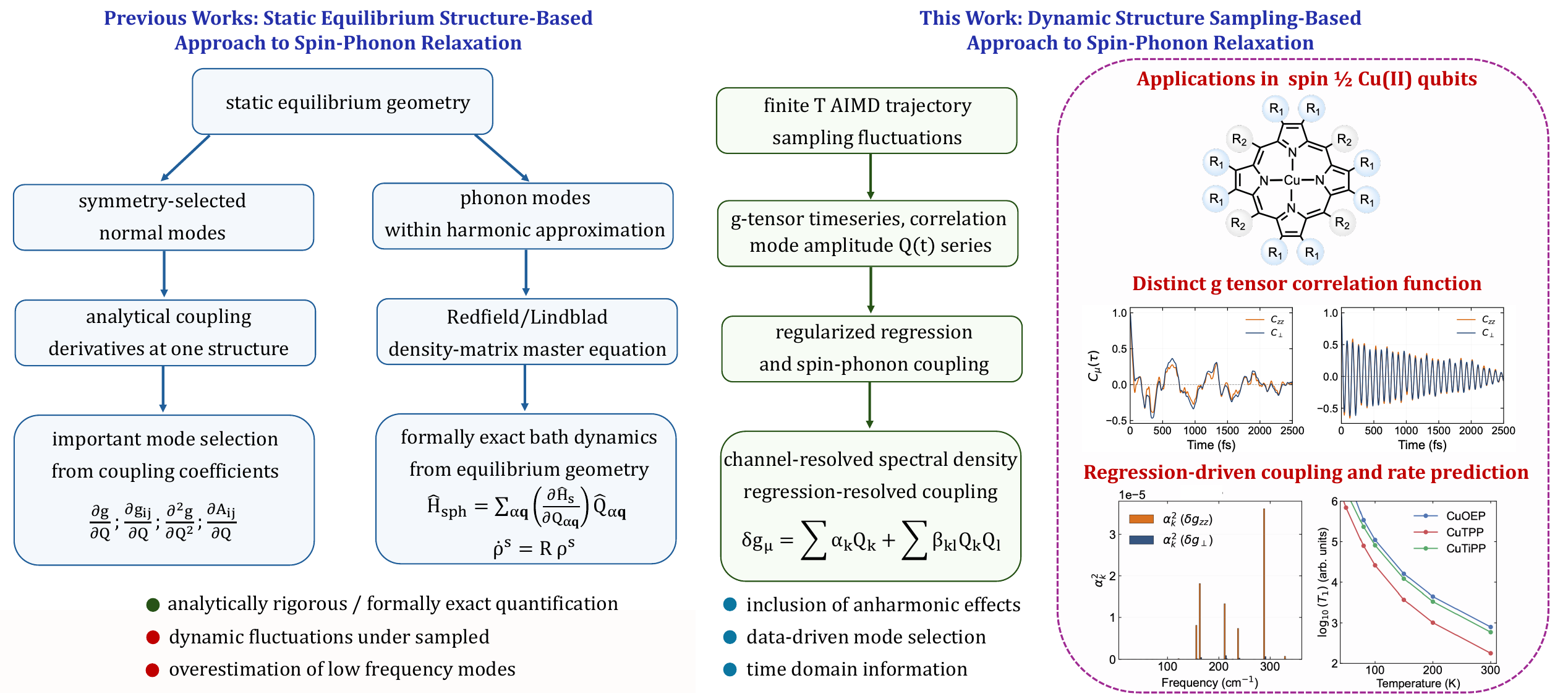}
\caption{Conceptual comparison of theoretical approaches to
spin--phonon relaxation in molecular qubits. Left panel: prior
approaches, based on symmetry-selected analytical coupling
derivatives at a static equilibrium geometry or on an open quantum
system density-matrix (Redfield/Lindblad) formulation propagated
from harmonic phonon modes. Right panel: this work, a
first-principles, regularized-regression (LASSO) approach that
calculates spin--phonon coupling constants directly from AIMD
trajectories. Applied to $S=\tfrac{1}{2}$ \ce{Cu^{II}} porphyrin
qubits, it explains prior experiments in terms of relative
$T_1$ relaxation rates and the predicted mode-frequency range
of dominant coupling.}
\label{fig:fig1}
\end{figure*}

Existing theoretical treatments of spin--phonon coupling have advanced considerably through open-quantum-system and master-equation formalisms, which make it possible to calculate mode-specific relaxation pathways (Fig.~\ref{fig:fig1}).~\cite{lunghi2019phonons,lunghi2022exact,briganti2025machine} In perturbative, normal-mode-based treatments, spin--phonon coupling derivatives are evaluated at a single optimized geometry and combined with a phonon bath to obtain relaxation rates.~\cite{lunghi2020limit,lunghi2019phonons,mirzoyan2020dynamic,kazmierczak2021impact,albino2019inorgchem,santanni2021inorgchem,Lee2026} These methods have been highly successful, but they are inherently tied to harmonic potentials and local derivatives around equilibrium, and they typically emphasize one or a few dominant modes rather than a fully mode-resolved, finite-temperature picture.~\cite{kazmierczak2024chemsci,garlatti2023natcomm} As a result, they can miss anharmonic effects and dynamic multi-mode contributions, especially for low-frequency motions. This limitation is particularly relevant here because low-frequency distortions are often strongly anharmonic, so harmonic normal-mode analysis may not adequately capture their role in relaxation.

Ab initio molecular dynamics (AIMD), by contrast, samples finite-temperature atomic displacements, implicitly incorporating dynamic inter-mode couplings and anharmonicity.~\cite{briganti2025machine,Sharma2020,egger2018advmater,yaffe2017prl,thomas2010sed,zhou2019prb,zhou2022prl} Yet AIMD introduces a different bottleneck: outside the perturbative normal-mode framework, there is no general analytical route for extracting the linear and higher-order spin--phonon coupling constants needed to connect finite-temperature fluctuations to direct and Raman relaxation channels.~\cite{briganti2025machine} In other words, AIMD naturally captures the relevant dynamics, but translating those fluctuations into physically interpretable, mode-resolved coupling constants remains nontrivial.

Here we address this problem using an automated, regression-driven framework (Fig.~\ref{fig:fig1}). We compute $g$-tensor time series along AIMD trajectories and project the normal-mode basis onto those trajectories to obtain mode amplitudes $Q_k(t)$ and bilinear mode--mode products $Q_k(t)Q_l(t)$. We then regress the channel-resolved fluctuations $\delta g_\mu(t)$, treated separately for each channel $\mu \in \{xx,yy,zz,\perp,\mathrm{iso}\}$, onto these linear and bilinear mode terms using least absolute shrinkage and selection operator (LASSO) regularization. This procedure yields sparse, physically interpretable linear ($\alpha_k$) and bilinear ($\beta_{kl}$) spin--phonon coupling coefficients for each $g$-tensor channel, thereby separating coupling orders. Although these coefficients are obtained from regression on sampled fluctuations rather than from analytical derivatives at a single equilibrium geometry, the approach is not restricted to harmonic, low-order expansions around one optimized structure and is therefore well suited to systems with substantial anharmonicity.

\begin{figure*}[t]
\centering
\includegraphics[width=0.9\textwidth]{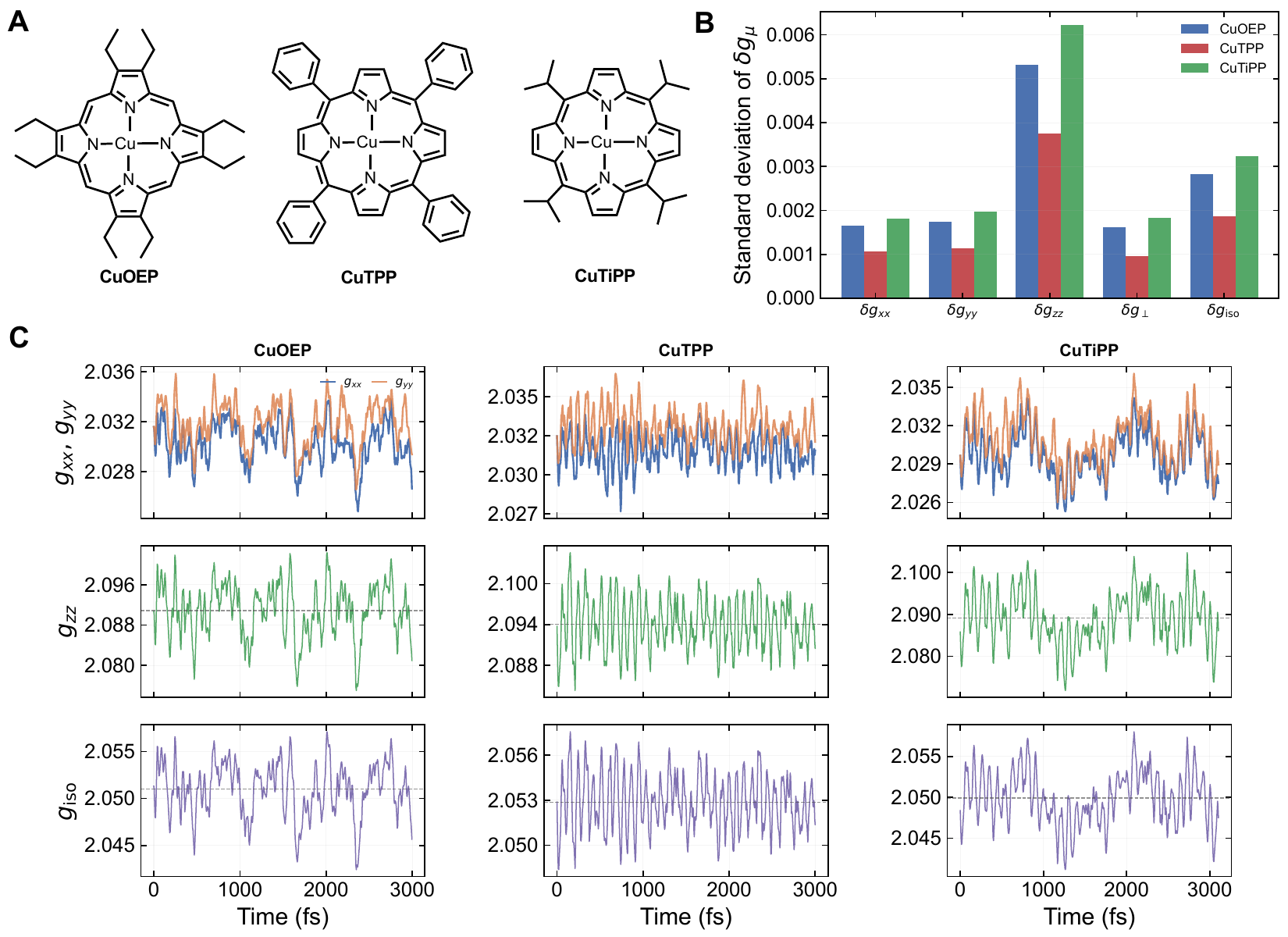}
\caption{$g$-tensor fluctuations from AIMD across the three Cu(II)
porphyrin systems. (A) Molecular structures of \ce{CuOEP},
\ce{CuTPP}, and \ce{CuTiPP}. (B) Standard deviation of $\delta g_\mu$
for each tensor channel ($xx$, $yy$, $zz$, $\perp$, iso), showing
the largest fluctuation amplitudes in $g_{zz}$ across all three
systems, with \ce{CuTiPP} slightly exceeding \ce{CuOEP} in most
channels and both consistently larger than \ce{CuTPP}. (C)
$g$-tensor time series ($g_{xx}$, $g_{yy}$, $g_{zz}$,
$g_{\mathrm{iso}}$) over the 3~ps AIMD trajectory for \ce{CuOEP},
\ce{CuTPP}, and \ce{CuTiPP}, illustrating the larger-amplitude
$g_{zz}$ fluctuations relative to the in-plane and isotropic
channels in all three systems.
}
\label{fig:fig2}
\end{figure*}

We apply this framework to CuOEP, CuTPP, and CuTiPP, a representative series of spin-$\tfrac{1}{2}$ \ce{Cu(II)} porphyrin qubits spanning different degrees of ruffling distortion (Fig.~\ref{fig:fig1}).~\cite{kazmierczak2024chemsci,lohaus2026jacs} The method recovers the key modes identified experimentally: rather than overemphasizing low-frequency distortions, it identifies bond-stretching modes in the 150--300~cm$^{-1}$ region as the dominant relaxation channels. While the spin spectral density $J_{\mu\mu}(\omega)$ reveals multiple frequency-resolved relaxation channels, regularized regression isolates the subset of modes that actually governs relaxation and corrects the tendency of spectral-density-only analysis to overestimate low-frequency contributions. We further show that this fluctuation-based approach captures physics inaccessible to a single frozen-phonon geometry: despite their structural similarity, the three qubits exhibit qualitatively distinct spin spectral densities that emerge only when finite-temperature fluctuations are sampled. Propagating the extracted coupling coefficients into $T_1(T)$ through both one-phonon (direct) and two-phonon (Raman) contributions reproduces the experimentally observed non-monotonic curvature in the relaxation profile and correctly predicts the relative relaxation rates among CuOEP, CuTPP, and CuTiPP.\cite{kazmierczak2024chemsci} Overall, these results establish a channel-resolved, coupling-order-separated route to spin--phonon analysis directly from AIMD fluctuations and provide a general framework for studying relaxation pathways in molecular spin qubits beyond the Cu(II) porphyrin family considered here.

\section{Results and Discussion}

\subsection{$g$-Tensor Time Series and Multi-Channel Fluctuations}
From AIMD trajectories of \ce{CuOEP}, \ce{CuTPP}, and \ce{CuTiPP} (Fig.~\ref{fig:fig2}A),
the $g$-tensor was evaluated at every 5~fs timestep (see Methods) to
quantify how thermal nuclear motion modulates the spin Hamiltonian.
The resulting time series show that these fluctuations
are strongly channel dependent and differ substantially in magnitude
across the three systems.
A nontrivial trend emerges in the overall fluctuation amplitude.
Although the degree of ruffling follows the structural order CuTiPP $>$
CuTPP $>$ CuOEP,~\cite{kazmierczak2024chemsci} the magnitude of the
$g$-tensor fluctuations instead follows CuTiPP $\sim$ CuOEP $>$ CuTPP (Fig.~\ref{fig:fig2}B and C).
While the high fluctuation amplitude of CuTiPP is consistent with its
greater ruffling distortion, the comparably large fluctuations
observed for CuOEP -- the least ruffled of the three systems -- show
that static structural distortion alone does not predict the
magnitude of thermally sampled $g$-tensor fluctuations. The
time-domain profiles further reveal clear differences in fluctuation
character: CuTPP exhibits relatively higher-frequency oscillations
than either CuOEP or CuTiPP (Fig.~\ref{fig:fig2}C). In all three systems, the out-of-plane
component dominates the response, with the standard deviation of
$\delta g_{zz}$ exceeding that of the in-plane components
$\delta g_{xx}$ and $\delta g_{yy}$ by a factor of 2--3 (Fig.~\ref{fig:fig2}B). Together, these results show that $g$-tensor
fluctuations are both anisotropic and system specific, underscoring
the importance of finite-temperature dynamics in shaping
spin-relaxation pathways even among closely related molecular
qubits.

\begin{figure*}[t]
\centering
\includegraphics[width=1\textwidth]{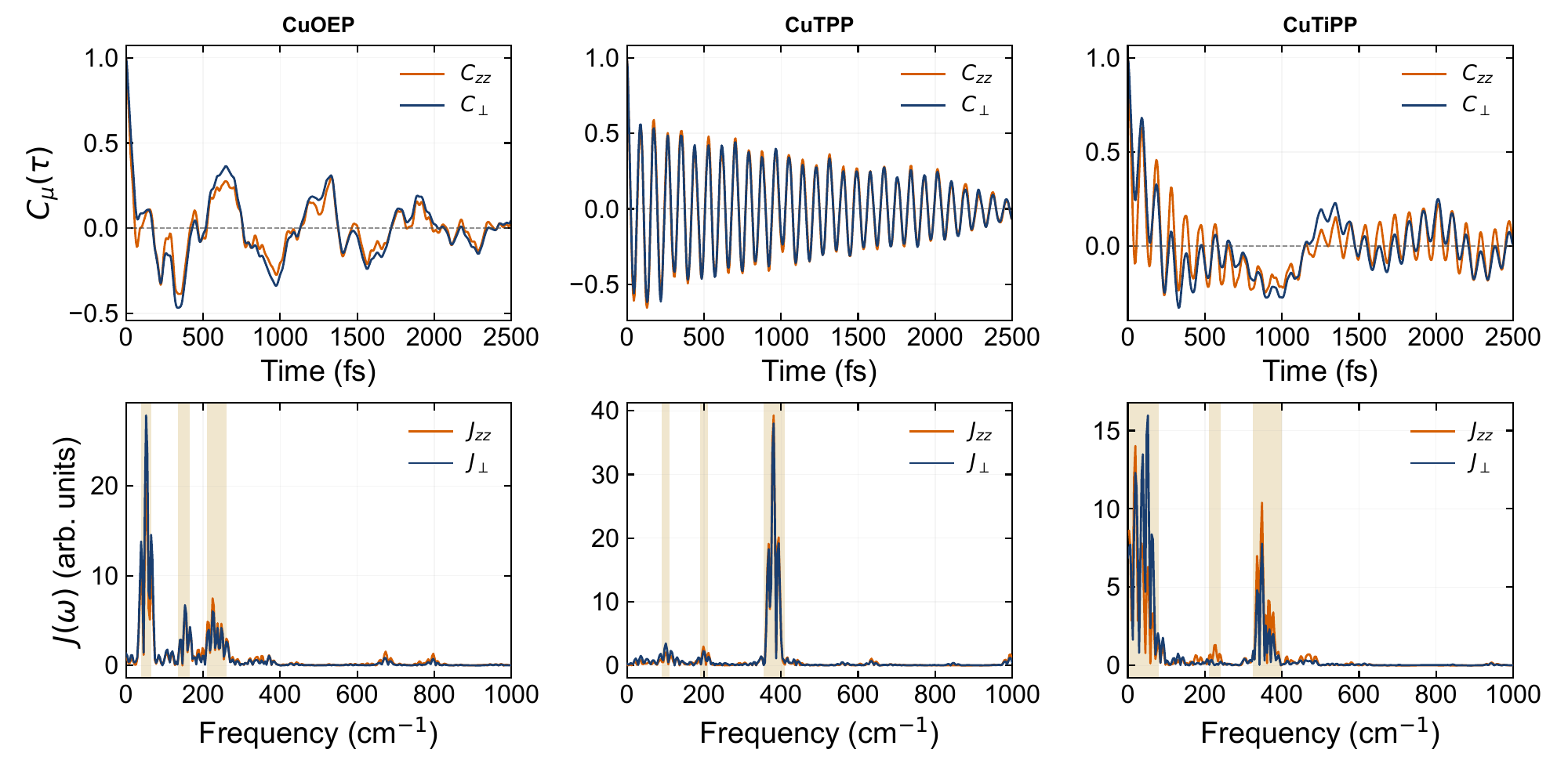}
\caption{Channel-resolved spin--phonon spectral density and
$g$-tensor autocorrelation for the three Cu(II) porphyrin systems.
Top row: normalized $g$-tensor autocorrelation functions $C_\mu(\tau)$
for the $zz$ and $\perp$ channels over the 2500~fs lag window.
\ce{CuOEP} and \ce{CuTiPP} exhibit slowly damped oscillatory decay
with pronounced channel splitting at longer lag times, while
\ce{CuTPP} shows rapid, closely overlapping oscillations that decay
fully within the trajectory window. Bottom row: corresponding
channel-resolved spectral densities $J_{zz}(\omega)$ and
$J_\perp(\omega)$. \ce{CuOEP} and \ce{CuTiPP} show multiple
low-to-mid-frequency peaks with appreciable channel splitting,
whereas \ce{CuTPP} is dominated by a single sharp high-frequency
peak with negligible low-frequency spectral weight. Shaded bands
indicate frequency ranges of dominant spectral weight discussed in
the text.
}
\label{fig:fig3}
\end{figure*}

\subsection{Tensor-Resolved Spin--Phonon Spectral Density}
To resolve both the amplitude and temporal character of the
$g$-tensor dynamics, we computed the channel-resolved spin--phonon
spectral density $J_\mu(\omega)$ for each tensor channel
$\mu \in \{xx, yy, zz, \perp, \mathrm{iso}\}$. For each channel, the
autocorrelation function of the $g$-tensor fluctuations
$\delta g_\mu(t)$ (defined in Methods) is
\begin{equation}
C_\mu(\tau) = \frac{\langle \delta g_\mu(t)\, \delta g_\mu(t+\tau)
\rangle_t}{\langle \delta g_\mu(t)^2 \rangle_t},
\label{eq:Ctau}
\end{equation}
where $\langle \cdot \rangle_t$ denotes averaging over trajectory
origins $t$. The corresponding spectral density is obtained via the
Wiener--Khintchine theorem~\cite{wiener1930harmonic,khintchine1934korrelationstheorie,berens1981jcp,thomas2013pccp} as the Fourier transform of $C_\mu(\tau)$,
\begin{equation}
J_\mu(\omega) = \int_{-\infty}^{\infty} C_\mu(\tau)\, w(\tau)\,
e^{-i\omega\tau}\, d\tau,
\label{eq:Jomega}
\end{equation}
where $w(\tau)$ is a Hanning window (see Methods). $J_\mu(\omega)$
therefore quantifies the frequency-resolved fluctuation power in
channel $\mu$. Resolving $J_{xx}(\omega)$, $J_{yy}(\omega)$, and
$J_{zz}(\omega)$ separately, rather than collapsing the dynamics into
a single isotropic spectral density, exposes channel-selective
spectral weight that would otherwise be obscured.

The autocorrelation functions and spectral densities reveal clear
differences among \ce{CuOEP}, \ce{CuTPP}, and \ce{CuTiPP} (Fig.~\ref{fig:fig3}).~\cite{kazmierczak2024chemsci} In \ce{CuOEP}, the autocorrelation
function decays relatively slowly, consistent with substantial
low-frequency content, and the spectral density is dominated by three
broad regions centered at 10--40~cm$^{-1}$, 170--190~cm$^{-1}$, and
200--230~cm$^{-1}$ (Fig.~\ref{fig:fig3}). By contrast, \ce{CuTPP} exhibits a more rapidly
oscillatory autocorrelation profile and predominantly higher-frequency
spectral weight, including pronounced features near 400~cm$^{-1}$,
consistent with the higher-frequency character already evident in its
$g$-tensor time series. \ce{CuTiPP} displays mixed behavior: its
autocorrelation function contains both slowly decaying and more
rapidly oscillating components, and its spectral density contains both
low- and high-frequency features (Fig.~\ref{fig:fig3}). Thus, from the perspective of
$g$-tensor dynamics, \ce{CuTiPP} lies between \ce{CuOEP} and
\ce{CuTPP}, rather than following a simple structural trend based on
static ruffling distortion alone. The intermediate behavior of CuTiPP is chemically reasonable. \ce{CuOEP} is
substituted with alkyl (ethyl) groups at the pyrrole
$\beta$-positions, leaving the meso positions unsubstituted, whereas
both \ce{CuTPP} and \ce{CuTiPP} carry substituents at the meso
positions -- aryl (phenyl) groups in \ce{CuTPP} and bulkier alkyl
(isopropyl) groups in \ce{CuTiPP}. \ce{CuTiPP} thus shares its
substitution position with \ce{CuTPP} but the alkyl
character of its substituents with \ce{CuOEP}. 

\begin{figure*}[t]
\centering
\includegraphics[width=1\textwidth]{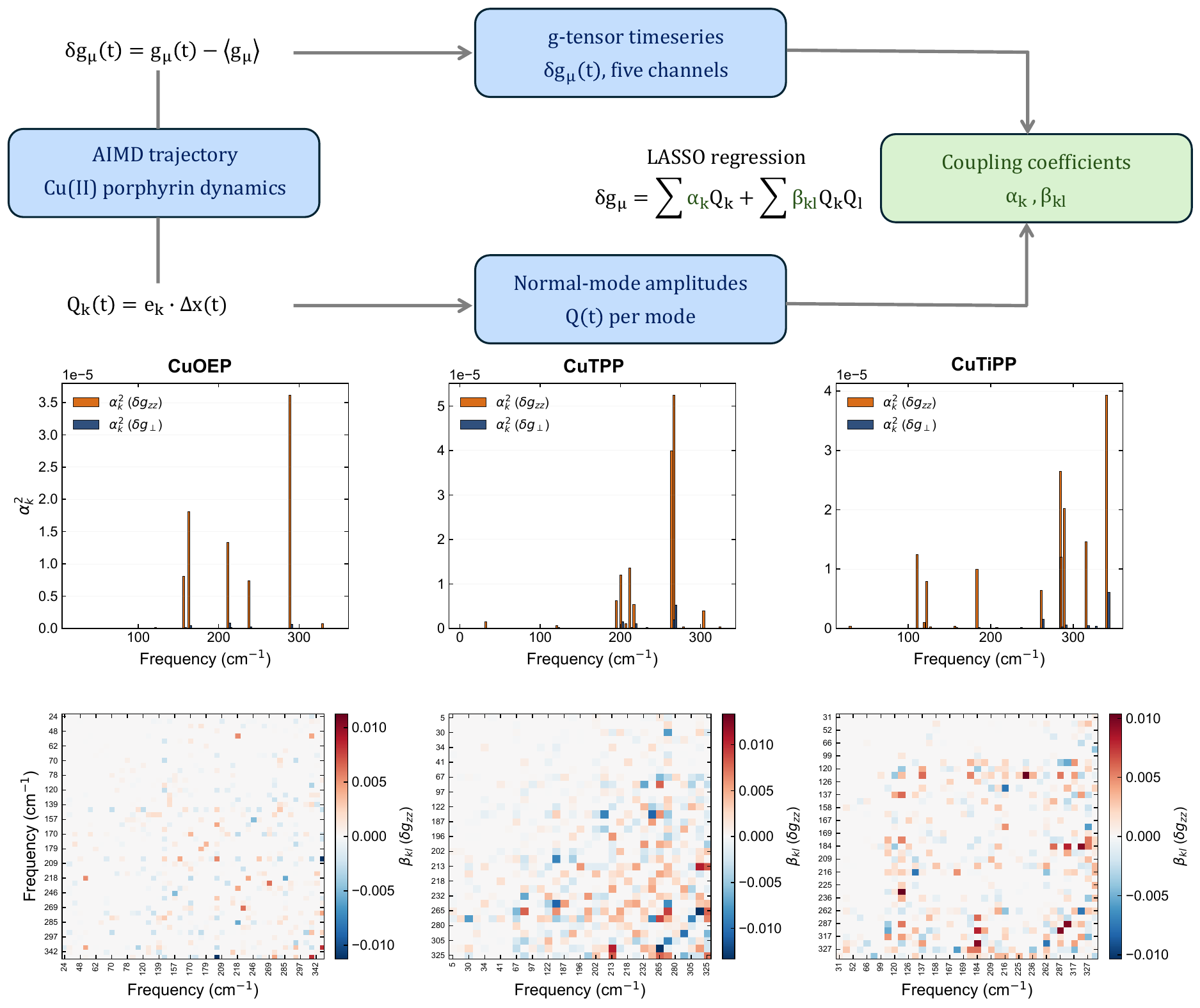}
\caption{LASSO regression workflow and channel-resolved coupling
coefficients. Top: schematic of the regression pipeline.
$g$-tensor fluctuations $\delta g_\mu(t)$ (from the AIMD trajectory)
and normal-mode amplitudes $Q_k(t)$ (projected mode displacements)
are related by LASSO regression [Eq.~\ref{eq:lasso_model_main}] to
yield the linear ($\alpha_k$) and bilinear ($\beta_{kl}$) coupling
coefficients. Middle row: regression-derived linear coupling
strengths $\alpha_k^2$ as a function of mode frequency for the $zz$
and $\perp$ channels in \ce{CuOEP}, \ce{CuTPP}, and \ce{CuTiPP}, showing
system-specific localization of dominant coupling to distinct
frequency ranges. Bottom row: bilinear coupling coefficients
$\beta_{kl}(\delta g_{zz})$ as a function of mode-pair frequency for
each system, showing the distribution of two-phonon coupling weight
across the mode spectrum.
}
\label{fig:fig4}
\end{figure*}

Overall, these tensor-resolved spectral densities provide direct
computational evidence that multiple spin channels are active in all three systems, and that
chemically related porphyrin qubits can exhibit distinct,
channel-selective spectral signatures despite their closely similar
core structures. We next connect these spectral features
to specific vibrational motions to identify which normal modes
dominate spin relaxation.

\subsection{LASSO Spin--Phonon Coupling: Linear ($\alpha_k$) and
            Bilinear ($\beta_{kl}$) Terms}
To isolate the intrinsic coupling efficiency of each normal mode, we project the atomic displacements ($\Delta \mathbf{x}$) from each AIMD trajectory frame
(defined in Methods) onto the normal-mode
eigenvectors $\mathbf{e}_k$, obtained by diagonalizing the
Hessian at the optimized geometry, to give the
time-dependent mode amplitude
\begin{equation}
  Q_k(t) = \mathbf{e}_k \cdot \Delta \mathbf{x}(t),
  \label{eq:mode_amplitude}
\end{equation}
for each retained normal mode $k$. The $g$-tensor fluctuations
$\delta g_\mu(t)$ are then regressed onto these mode amplitudes and
their pairwise products using cross-validated LASSO regression. For
each tensor channel $\mu$, the regression model is
\begin{equation}
  \delta g_\mu(t) = \sum_k \alpha_{k,\mu}\,Q_k(t)
  + \sum_{k \le l} \beta_{kl,\mu}\,Q_k(t)\,Q_l(t) + \varepsilon(t),
  \label{eq:lasso_model_main}
\end{equation}
where the linear coefficients $\alpha_{k,\mu}$ capture first-order
(one-phonon) spin--phonon coupling and the bilinear coefficients
$\beta_{kl,\mu}$ capture second-order (two-phonon, Raman-active)
coupling between mode pairs (Top panel in Fig.~\ref{fig:fig4}), and $\varepsilon(t)$ is the residual capturing higher-order and
unmodeled contributions (Table I, III, V). We note that the vibrational modes discussed here are molecular
(rather than lattice) modes. Nonetheless, to remain consistent with
the established literature, we refer to the corresponding couplings
as spin--phonon couplings.

\begin{figure*}[t]
\centering
\includegraphics[width=1\textwidth]{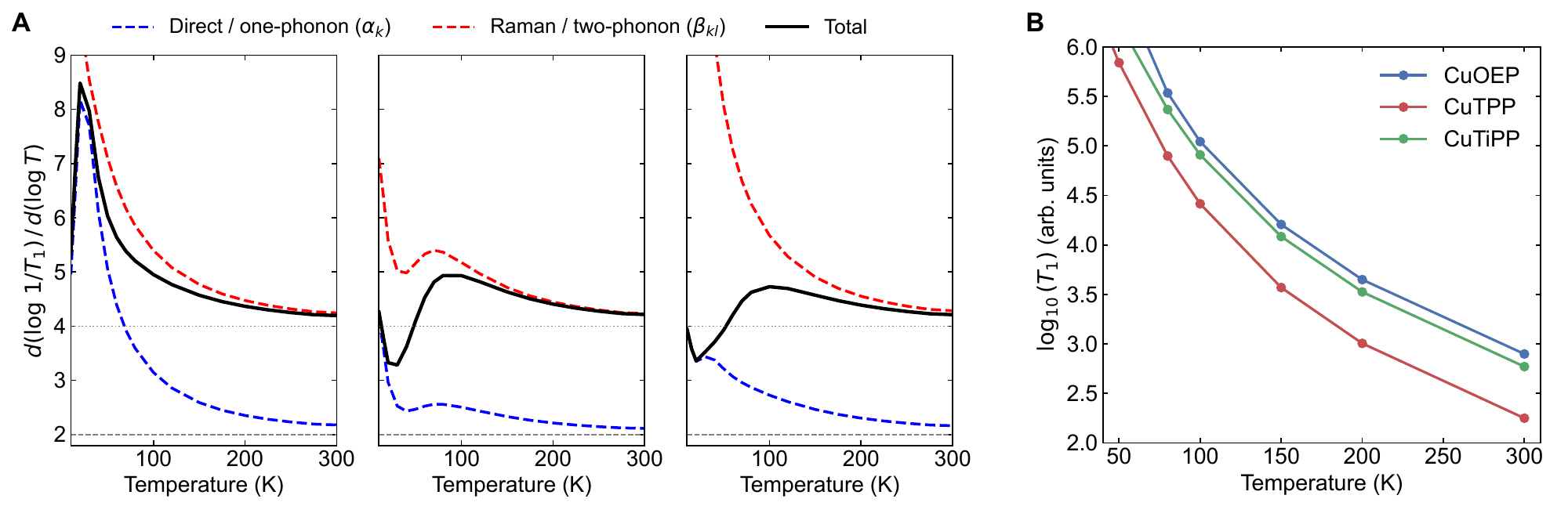}
\caption{Temperature dependence of the spin--lattice relaxation
rate across the Cu(II) porphyrin series. (A) Local power-law
exponent $d(\log 1/T_1)/d(\log T)$ as a function of temperature for
\ce{CuOEP}, \ce{CuTPP}, and \ce{CuTiPP} (left to right), decomposed
into the direct, one-phonon contribution ($\alpha_k$, blue dashed),
the Raman, two-phonon contribution ($\beta_{kl}$, red dashed), and
the total rate (black solid). Horizontal reference lines mark slopes
of 2 and 4, the limiting power-law behaviors expected for direct and
Raman relaxation, respectively. (B) $\log_{10}(T_1)$ versus
temperature for all three systems (\ce{CuOEP}, blue; \ce{CuTPP},
red; \ce{CuTiPP}, green), computed from the same LASSO-derived
coupling coefficients.}
\label{fig:fig5}
\end{figure*}

\subsubsection{Mode-Resolved Coupling from LASSO Regression}
We report LASSO-regressed spin--phonon coupling coefficients for
\ce{CuOEP}, \ce{CuTPP}, and \ce{CuTiPP} (Fig.~\ref{fig:fig4}, middle and bottom panel);
regression and cross-validation details are given in the Supporting
Information (page S3-S9). In all three systems, multiple normal modes are
identified as significant for the $g$-tensor channels (Fig.~\ref{fig:fig4}). Consistent
with experimental resonance Raman and neutron assignments,\cite{kazmierczak2024chemsci,lohaus2026jacs} dominant
coupling in all cases falls in the 100--300~cm$^{-1}$ range rather
than below 100~cm$^{-1}$.

For \ce{CuOEP}, the $\delta g_{zz}$ channel is dominated by a single
mode near 289~cm$^{-1}$, with secondary contributions near 164, 213,
and 238~cm$^{-1}$, close to the experimentally assigned
271~cm$^{-1}$ local mode.\cite{kazmierczak2024chemsci} By contrast,
the $\delta g_\perp$ channel shows no single dominant mode: coupling
is spread thinly across the same 150--300~cm$^{-1}$ region at
substantially smaller magnitude than $\delta g_{zz}$, showing that
the mode structure driving $g_{zz}$ fluctuations does not carry over
directly to $g_\perp$. Low-frequency ruffling modes below
100~cm$^{-1}$ carry negligible coupling in both channels despite
dominating the spectral density $J_\mu(\omega)$, illustrating why large
ruffling amplitude does not translate into ruffling-driven
relaxation.\cite{kazmierczak2024chemsci,lohaus2026jacs}

In \ce{CuTPP}, dominant $\alpha_k$ coefficients for $\delta g_{zz}$
cluster tightly in the 265--268~cm$^{-1}$ range. Additionally,
multiple modes near 200~cm$^{-1}$ are also found to be important.
This is consistent with the experimental resonance Raman assignment,
which identifies lower-energy modes for \ce{CuTPP} than for
\ce{CuOEP}.\cite{kazmierczak2024chemsci} \ce{CuTiPP} shows a distinct
pattern from both other systems: dominant $\delta g_{zz}$ coupling is
concentrated at the high-frequency end of the retained window, near
300~cm$^{-1}$ and above, alongside a secondary lower-frequency
cluster near 112--123~cm$^{-1}$ (Figure~2C).

Comparing the frequency region where the dominant, high-coupling
modes are concentrated across the three systems, this concentration
shifts progressively higher in frequency in the order \ce{CuTPP} $<$
\ce{CuOEP} $\sim$ \ce{CuTiPP}. This ordering is consistent with the
resonance Raman mode assignments used to explain relaxation rates
experimentally.\cite{kazmierczak2024chemsci} 

Beyond these linear coefficients, the bilinear terms
$\beta_{kl,\mu}$ extracted from the same regression give direct
access to the two-phonon coupling network (Figure~2, bottom row). In
\ce{CuOEP}, resolvable $\beta_{kl}$ pairs are sparse and of small
magnitude. \ce{CuTPP} shows a markedly denser bilinear coupling
network, with dominant pairs clustering within the
200--330~cm$^{-1}$ bond-stretch regime itself, including pronounced
self-coupling near $\omega_k \approx \omega_l \approx
265$--$280$~cm$^{-1}$. \ce{CuTiPP} shows the densest bilinear
network of the three systems, with strong cross-coupling both within
the 260--330~cm$^{-1}$ region and between this region and
lower-frequency modes near 120--190~cm$^{-1}$, consistent with its
broader spread of significant linear-coupling modes.

The linear and bilinear coefficients resolve the ambiguity
left by the spectral density analysis. Spin--phonon coupling across
all three systems concentrates in bond-stretch modes above
100~cm$^{-1}$, consistent with resonance Raman and neutron
scattering assignments, while the low-frequency ruffling modes that
dominate $J(\omega)$ contribute only weakly to true coupling.~\cite{kazmierczak2024chemsci,lohaus2026jacs} This
distinction between fluctuation amplitude and coupling strength is
the central result enabling the $T_1(T)$ predictions in the
following section.

\subsection{$T_1$ Temperature Dependence: Direct vs.\ Raman
            Decomposition and Relative Relaxation Rates}

The LASSO-derived coupling coefficients $\alpha_{k,\mu}$ and
$\beta_{kl,\mu}$ were propagated into temperature-dependent
relaxation rates through one-phonon (direct) and two-phonon (Raman)
contributions~\cite{lunghi2023spinphonon,mirzoyan2020dynamic,vanvleck1940pr},
\begin{align}
  \left(\frac{1}{T_1}\right)_{\!\mathrm{direct}}
  &= A\sum_k \bar\alpha_k^2 \,
     \frac{e^{\hbar\omega_k/k_BT}}{(e^{\hbar\omega_k/k_BT}-1)^2},
  \label{eq:T1_direct_main}\\
  \left(\frac{1}{T_1}\right)_{\!\mathrm{Raman}}
  &= A_2\sum_{k,l} \bar\beta_{kl}^2 \, n_k(n_k+1)\,n_l(n_l+1),
  \label{eq:T1_raman_main}
\end{align}
where $\bar\alpha_k^2$ and $\bar\beta_{kl}^2$ are the LASSO
coefficients averaged over the $xx$, $yy$, and $zz$ $g$-tensor
components and $n_k = (e^{\hbar\omega_k/k_BT}-1)^{-1}$ is the
Bose--Einstein occupation number. $A$ and $A_2$ are system- and temperature-independent scale factors,
set to unity here since the framework reports channel-resolved
\emph{relative} contributions to $1/T_1$ in arbitrary units (see page S3-S4). The
temperature dependence and channel decomposition are therefore fully
determined by $\bar\alpha_k^2$ and $\bar\beta_{kl}^2$, so the
bond-stretch modes identified in Figure \ref{fig:fig4} as dominant in
$\alpha_k$ and $\beta_{kl}$ also dominate the resulting rate.

The direct and Raman contributions exhibit distinct temperature
dependence, and their relative weight governs the curvature of
$1/T_1(T)$ in each system (Fig. \ref{fig:fig5}A and Table II, IV, VI). In \ce{CuOEP}, the bilinear
(Raman) term requires simultaneous two-phonon population and becomes
appreciable only once the dominant 200--290~cm$^{-1}$ modes are
thermally populated ($\sim$40--80~K); below this range direct
relaxation dominates, while above it the Raman contribution
overtakes and the total rate curvature reflects this
direct-to-Raman crossover. In \ce{CuTPP}, the direct process
contributes negligibly across the entire temperature range, so its rate is governed almost
entirely by the Raman channel, with curvature instead arising from
sequential activation of its 122~cm$^{-1}$, $\sim$200~cm$^{-1}$, and
$\sim$260~cm$^{-1}$ mode families. \ce{CuTiPP} shows an intermediate
profile, consistent with its mixed low-- and high--frequency
coupling structure identified in the LASSO analysis (Fig. \ref{fig:fig4}).

Propagating these rates to absolute $T_1(T)$ values reproduces the
relative relaxation trend observed experimentally (Fig. \ref{fig:fig5}B):
\ce{CuTPP} relaxes fastest across the entire measured temperature
range, consistent with its dense bilinear coupling network and
correspondingly large Raman-to-direct ratio. \ce{CuOEP} and
\ce{CuTiPP} relax more slowly and track closely with
one another across the same range, with \ce{CuTiPP} consistently
falling slightly below \ce{CuOEP}, indicating marginally faster
relaxation. This three-way ordering -- \ce{CuTPP} fastest, followed
closely by \ce{CuTiPP} and then \ce{CuOEP} -- matches the
experimental $T_1$ trend directly.

\subsection{Outlook and Conclusions}
Our protocol introduces an alternative approach based on regularized
regression to calculate spin--phonon coupling constants and the
$T_1$ vs.\ $T$ dependence, in contrast to open-quantum-system
Redfield approaches based on the master
equation.~\cite{lunghi2023spinphonon} Our method does not currently provide the analytical rigor of these
treatments; it does, however, offer a systematic, data-driven route
to the coupling constants and demonstrates that time-domain
information is essential to include. Further, the current framework is based on a 3~ps trajectory;
tracking nanosecond-to-microsecond-scale processes will require
substantially longer trajectories, which lie beyond the reach of
AIMD and demand further methodological development. Additionally,
only the $g$-tensor channel for a spin-1/2 qubit is demonstrated
here, and more work is needed to extend the framework to other
systems, with potential involvement of processes such as spin-orbit
coupling, Orbach relaxation, and spin-spin interactions.

Overall, we have introduced a first-principles pipeline that resolves spin--phonon coupling in molecular qubits by $g$-tensor channel and by coupling order, separating linear (one-phonon) and bilinear (two-phonon, Raman) contributions. These channel-resolved coefficients are extracted via LASSO regression on AIMD-derived mode displacements and propagated into $T_1(T)$ without additional fitting to the relaxation shape itself. Moreover, we provide insights into the time-domain fluctuations of $g$-tensor channels, which show distinct behavior even within similar chemical structures. Our results capture the qualitative relaxation rate profile and relative ordering across systems, along with correct identification of the relevant mode range. This resolves the low-frequency overestimation issue highlighted by resonance Raman and inelastic neutron scattering experiments. The results show promise for extension to other classes of molecular qubits and highlight the methodological advancements needed to improve the state of the proposed approach.

\section{Methods}
We studied \ce{CuOEP} (copper(II) octaethylporphyrin), \ce{CuTPP}
(copper(II) tetraphenylporphyrin), and \ce{CuTiPP} (copper(II)
tetra(isopropyl)porphyrin), all square-planar complexes for which experimental $T_1(T)$ data
are available.\cite{kazmierczak2024chemsci,lohaus2026jacs} AIMD
trajectories were generated in ORCA 6.1.0~\cite{neese2020orca} at the
TPSS-D3BJ/def2-SVP level.~\cite{tao2003tpss,weigend2005def2,Grimme2010_D3} Initial velocities were sampled at 300~K, and
trajectories were propagated in the canonical ensemble using a
Berendsen thermostat (target 300~K, time constant 10~fs) with a
0.5~fs integration timestep, for a total production length of
$\approx$3~ps (6000 frames) for all three systems. $g$-tensor
snapshots were computed every 10 AIMD steps (5~fs sampling) at the
PBE/def2-SVP level (def2-TZVP on Cu and coordinating
N),\cite{perdew1996pbe} yielding $\approx$600 snapshots per
trajectory. Harmonic frequencies at the same level of theory
provided the normal-mode basis. Throughout this work we use ``phonon'' in the conventional spin-phonon coupling sense to denote the vibrational degrees of freedom that modulate the spin Hamiltonian; because our systems are treated as isolated molecules rather than periodic solids, these are more precisely molecular vibrational normal modes extracted from the AIMD trajectories, and we compute vibrational modes accordingly rather than crystal phonons with associated dispersion.

Mass-weighted atomic displacements $\Delta \mathbf{x}_i(t) =
\sqrt{m_i}\,\Delta \mathbf{r}_i(t)$ from the equilibrium geometry
were computed for each trajectory frame after removing
center-of-mass drift and rigid-body rotation, and projected onto the
mass-weighted normal-mode eigenvectors to give the mode amplitudes
$Q_k(t)$ of Eq.~\ref{eq:mode_amplitude}. Modes corresponding to
translation, rotation, and imaginary frequencies were excluded prior
to projection.

$g$-tensor fluctuations $\delta g_\mu(t) = g_\mu(t) - \langle g_\mu
\rangle$ were computed for each tensor channel $\mu \in \{xx, yy,
zz, \perp, \mathrm{iso}\}$ from the AIMD trajectory, and the
corresponding autocorrelation and spectral density were obtained as
in Eqs.~\ref{eq:Ctau} and \ref{eq:Jomega}. The Hanning window
$w(\tau)$ was applied with a maximum lag of 2500~fs to suppress spectral leakage while retaining
sensitivity to slowly damped low-frequency oscillations.

The regression model of Eq.~\ref{eq:lasso_model_main} was fit
independently for each tensor channel $\mu$ and each system using
cross-validated LASSO regression (scikit-learn
LassoCV~\cite{pedregosa2011scikit}), with features restricted to
modes below 350~cm$^{-1}$ across all three systems (Table I, III, IV).

The LASSO-derived coefficients $\alpha_{k,\mu}$ and $\beta_{kl,\mu}$
were propagated into temperature-dependent relaxation rates through
the one-phonon (direct) and two-phonon (Raman) expressions of
Eqs.~\ref{eq:T1_direct_main} and \ref{eq:T1_raman_main}, with
$\bar\alpha_k^2$ and $\bar\beta_{kl}^2$ averaged over the $xx$,
$yy$, and $zz$ $g$-tensor components (see page S5).

\section{Author Information}

\textbf{Corresponding Author}

\textbf{Sayan Banerjee} — Department of Chemistry, University of Tennessee, Knoxville, Tennessee 37996, United States; orcid.org/0000-0002-8586-9236\\
\textbf{Email}: sbanerjee@utk.edu\\







\section{Supporting Information}

\subsection{AIMD Protocol}

Each system studied is a Cu(II) porphyrin-type molecular qubit ($S = \tfrac{1}{2}$, \ce{Cu^{II}}, $d^9$) with experimentally measured $T_1(T)$ data available for comparison.\cite{kazmierczak2024chemsci,lohaus2026jacs} Ab initio molecular dynamics trajectories were generated in ORCA 6.1.0 using the TPSS functional with a def2-SVP basis set, propagated in the NVT ensemble at $T = 300$~K with a timestep of $\Delta t = 0.5$~fs.\cite{neese2020orca} $g$-tensor snapshots were computed using the ORCA EPR module at the PBE/def2-SVP level, with def2-TZVP applied to the metal center and the coordinating nitrogen atoms, extracted every ten AIMD steps to yield a $g$-tensor sampling interval of $\Delta t_g = 5$~fs. A harmonic frequency calculation, performed at the same level of theory following geometry optimization, provided the reference normal-mode basis. Six rigid-body translational and rotational modes, together with any residual imaginary modes, were excluded from the projection basis.

\subsection{Mode Projection}

Prior to mode projection, each trajectory frame was corrected for center-of-mass drift and aligned to the reference structure (frame 0) via singular value decomposition. The atomic displacement at each frame was then projected onto each mass-weighted normal mode according to

\begin{equation}
  Q_k(t) = \sum_{i=1}^{3N} \sqrt{m_i}\,
           \bigl[r_i(t) - r_i^{(0)}\bigr]\,
           e_{k,i} ,
  \label{eq:mode_proj_si}
\end{equation}

where $e_{k,i}$ is the $i$-th Cartesian component of the $k$-th mass-weighted normal-mode eigenvector and $r_i^{(0)}$ is the aligned reference coordinate. Static (equal-time) coupling between modes was assessed via the Pearson correlation matrix,

\subsection{Channel-Resolved $g$-Tensor Spectral Density}

For each principal $g$-tensor component $\mu \in \{xx, yy, zz\}$, the fluctuation about its trajectory mean was defined as

\begin{equation}
  \delta g_\mu(t) = g_\mu(t) - \langle g_\mu \rangle ,
  \label{eq:dg_si}
\end{equation}

with the transverse and isotropic channels constructed as

\begin{align}
  \delta g_\perp &= \tfrac{1}{2}(\delta g_{xx} + \delta g_{yy}) ,
  \label{eq:gperp_si}\\
  \delta g_\mathrm{iso} &= \tfrac{1}{3}
  (\delta g_{xx}+\delta g_{yy}+\delta g_{zz}) .
  \label{eq:giso_si}
\end{align}

The normalized autocorrelation of each channel, evaluated at discrete lags $\tau = n\,\Delta t_g$, is given by

\begin{equation}
  C_\mu(\tau) =
  \frac{\langle \delta g_\mu(t+\tau)\,\delta g_\mu(t)\rangle}
       {\langle \delta g_\mu^2 \rangle} .
  \label{eq:autocorr_si}
\end{equation}

To obtain the spin--phonon spectral density, $C_\mu(\tau)$ was multiplied by a Hanning window $w(\tau)$, zero-padded to a length $N_\mathrm{fft} = 4\times 2^{\lceil \log_2 n\rceil}$ (with $n$ the number of retained lag points), Fourier-transformed, and the spectral density taken as the magnitude of the real part of the resulting transform,

\begin{equation}
  J_\mu(\omega) = \left|\,\operatorname{Re}
  \!\left[\sum_{\tau} C_\mu(\tau)\,w(\tau)\,
  e^{i\omega\tau}\right]\right| .
  \label{eq:Jomega_si}
\end{equation}

The autocorrelation lag window ($\tau_\mathrm{max} = n\,\Delta t_g$) was chosen for each system to be long enough to capture the slowest-damped low-frequency oscillation while retaining sufficient statistical averaging at long lag, which in practice required that at least approximately 15\% of the total number of $g$-tensor snapshots remain outside the lag window. It is important to note that $J_\mu(\omega)$ reflects the amplitude times the coherence time of each mode's contribution to $\delta g_\mu$, and does not by itself equal the spin--phonon coupling efficiency: a large-amplitude, long-lived low-frequency mode can dominate $J_\mu(\omega)$ even when its intrinsic per-displacement coupling constant is small. This distinction is resolved explicitly by the regression analysis described in the following section.

\subsection{LASSO Spin--Phonon Regression}

For each $g$-tensor channel $\mu \in \{xx,yy,zz,\perp,\mathrm{iso}\}$, the fluctuation time series was modeled as a sum of linear and bilinear contributions from the low-frequency normal modes,

\begin{equation}
  \delta g_\mu(t) = \sum_{k=1}^{n_\mathrm{low}} \alpha_{k,\mu}\,Q_k(t)
  + \sum_{k \le l} \beta_{kl,\mu}\,Q_k(t)\,Q_l(t)
  + \varepsilon(t) ,
  \label{eq:lasso_model_si}
\end{equation}

where the linear coefficients $\alpha_{k,\mu}$ represent first-order (one-phonon) spin--phonon coupling and the quadratic coefficients $\beta_{kl,\mu}$, which include the diagonal terms $k=l$, represent bilinear (two-phonon, Raman) coupling. A single frequency cutoff, $\Omega_c = \SI{350}{\per\cm}$, was applied identically across all systems to define the set of $n_\mathrm{low}$ modes retained in the regression, so that no system-specific tuning enters the feature construction. The resulting feature matrix has dimensions $N_\mathrm{snap} \times \left(n_\mathrm{low} + \binom{n_\mathrm{low}+1}{2}\right)$, combining the linear mode displacements with all pairwise (and self-paired) products.

Each feature column $j$ was standardized to zero mean and unit variance prior to fitting,

\begin{equation}
  X^{(\mathrm{sc})}_{\cdot j} =
  \frac{X_{\cdot j} - \mu_j}{\sigma_j} ,
  \label{eq:standardize_si}
\end{equation}

and the fitted coefficients were subsequently converted back to physical units by dividing elementwise by the corresponding feature scale factor,

\begin{equation}
  w^{(\mathrm{phys})}_j = \frac{w_j}{\sigma_j} ,
  \label{eq:unscale_si}
\end{equation}

with the set $\{\alpha_{k,\mu}\}$ recovered from the first $n_\mathrm{low}$ entries of $w^{(\mathrm{phys})}$ and $\{\beta_{kl,\mu}\}$ from the remaining $\binom{n_\mathrm{low}+1}{2}$ entries. Regularization was performed using scikit-learn's LassoCV implementation,~\cite{pedregosa2011scikit} which minimizes

\begin{equation}
  \frac{1}{2N_\mathrm{snap}}
  \bigl\lVert \delta g_\mu - X^{(\mathrm{sc})} w \bigr\rVert_2^2
  + \lambda \lVert w \rVert_1
  \label{eq:lasso_obj_si}
\end{equation}

over a grid of twenty candidate $\lambda$ values, with the optimal $\lambda$ selected by three-fold cross-validation.






Fit quality was assessed via the coefficient of determination,

\begin{equation}
  R^2 = 1 - \frac{\sum_t \bigl(\delta g_\mu(t) - \widehat{\delta g}_\mu(t)\bigr)^2}
                  {\sum_t \bigl(\delta g_\mu(t) - \overline{\delta g_\mu}\bigr)^2} ,
  \label{eq:r2_si}
\end{equation}

reported under the primary shuffled-cross-validation scheme. Physically, $\alpha_{k,\mu}^2$ represents the squared linear coupling coefficient, corresponding to the spin--phonon coupling efficiency per unit normal-mode displacement for channel $\mu$, and is the quantity used to rank dominant modes -- in contrast to $J_\mu(\omega)$, which reflects thermally weighted amplitude and coherence time rather than intrinsic coupling strength, as discussed above. Analogously, $\beta_{kl,\mu}^2$ represents the corresponding bilinear coupling strength for mode pair $(k,l)$.

\subsection{$T_1$ Temperature Dependence}

The one-phonon (direct) relaxation rate was computed following an established formalism for channel-averaged direct-process relaxation,\cite{mirzoyan2020dynamic,lunghi2023spinphonon}

\begin{equation}
  \left(\frac{1}{T_1}\right)_\mathrm{direct}
  = A\sum_k
  \frac{\alpha_{k,xx}^2+\alpha_{k,yy}^2+\alpha_{k,zz}^2}{3}
  \cdot
  \frac{e^{\hbar\omega_k/k_BT}}
       {(e^{\hbar\omega_k/k_BT}-1)^2} ,
  \label{eq:T1_direct_si}
\end{equation}

while the two-phonon (Raman) relaxation rate was computed as


\begin{widetext}
\begin{equation*}
  \left(\frac{1}{T_1}\right)_\mathrm{Raman}
  = A_2\sum_{k,l}
  \frac{\beta_{kl,xx}^2+\beta_{kl,yy}^2+\beta_{kl,zz}^2}{3}
  \cdot n_k(n_k+1)\cdot n_l(n_l+1) ,
  \label{eq:T1_raman_si}
\end{equation*}
\end{widetext}

where $n_k = (e^{\hbar\omega_k/k_BT}-1)^{-1}$ is the Bose--Einstein occupation number for mode $k$. The total relaxation rate is the sum of the two contributions,

\begin{equation}
  \left(\frac{1}{T_1}\right)_\mathrm{total}
  = \left(\frac{1}{T_1}\right)_\mathrm{direct}
  + \left(\frac{1}{T_1}\right)_\mathrm{Raman} .
  \label{eq:T1_total_si}
\end{equation}

Both $A$ and $A_2$ are treated as free multiplicative prefactors. Unless a given rate curve is explicitly calibrated against a known experimental $T_1$ value at a reference temperature, all reported rates are expressed in arbitrary units and are used solely for relative comparisons of shape and ordering across systems and channels.

The local power-law exponent characterizing the temperature dependence of the relaxation rate, an experimentally accessible observable, was defined as

\begin{equation}
  \eta(T) = \frac{\mathrm{d}\ln(1/T_1)}{\mathrm{d}\ln T} ,
  \label{eq:slope_si}
\end{equation}

and evaluated numerically using central finite differences on a logarithmic temperature grid. 






\section{Tables for LASSO Regression and $T_1$ Results}

\subsection{Fit Quality Summary for CuOEP}
Table~\ref{tab:lasso_fit_quality_cuoep} summarizes the LASSO fit quality across all five $g$-tensor channels. The shuffled-cross-validation $R^2$ values are uniformly high ($>0.98$).

\subsection{Temperature Dependence of the Relaxation Rate for CuOEP}
Table~\ref{tab:T1_rates_cuoep} reports the direct, Raman, and total relaxation rates on the temperature grid used for the $T_1(T)$ analysis, together with the local power-law exponent $\eta(T)$ defined in Eq.~\ref{eq:slope_si}. Rates are reported in arbitrary units, as no experimental calibration temperature was applied.

\subsection{Fit Quality Summary for CuTPP}
Table~\ref{tab:lasso_fit_quality_cutpp} summarizes the LASSO fit quality across all five $g$-tensor channels. The shuffled-cross-validation $R^2$ values are uniformly high ($>0.97$).

\subsection{Temperature Dependence of the Relaxation Rate for CuTPP}
Table~\ref{tab:T1_rates_cutpp} reports the direct, Raman, and total relaxation rates on the temperature grid used for the $T_1(T)$ analysis, together with the local power-law exponent $\eta(T)$ defined in Eq.~\ref{eq:slope_si}. Rates are reported in arbitrary units, as no experimental calibration temperature was applied.

\subsection{Fit Quality Summary for CuTiPP}

Table~\ref{tab:lasso_fit_quality} summarizes the LASSO fit quality across all five $g$-tensor channels. The shuffled-cross-validation $R^2$ values are uniformly high ($>0.98$).

\subsection{Temperature Dependence of the Relaxation Rate for CuTiPP}

Table~\ref{tab:T1_rates} reports the direct, Raman, and total relaxation rates on the temperature grid used for the $T_1(T)$ analysis, together with the local power-law exponent $\eta(T)$ defined in Eq.~\ref{eq:slope_si}. Rates are reported in arbitrary units, as no experimental calibration temperature was applied.

\begin{table*}[htbp]
\centering
\caption{LASSO fit-quality summary for all five $g$-tensor channels. $R^2_\mathrm{shuf}$ denotes the coefficient of determination under shuffled cross-validation. $n_\alpha$ and $n_\beta$ report the number of nonzero linear and bilinear coefficients out of the total available ($n_\mathrm{low}=50$ linear features; 1275 bilinear features).}
\label{tab:lasso_fit_quality_cuoep}
\begin{tabular}{lcccccc}
\toprule
Channel & $\lambda_\mathrm{best}$ & $R^2_\mathrm{shuf}$ & $n_\alpha$ (/50) & $n_\beta$ (/1275) & $\sigma_\mathrm{signal}$ & Expl.\ frac.\ \\
\midrule
$xx$        & \num{7.606e-7} & 0.9863 & 10 & 228 & \num{1.657e-3} & 0.883 \\
$yy$        & \num{7.895e-7} & 0.9871 &  6 & 207 & \num{1.736e-3} & 0.886 \\
$zz$        & \num{2.465e-6} & 0.9855 &  7 & 226 & \num{5.318e-3} & 0.880 \\
$\perp$     & \num{7.751e-7} & 0.9892 & 10 & 209 & \num{1.614e-3} & 0.896 \\
$\mathrm{iso}$ & \num{1.338e-6} & 0.9870 &  9 & 216 & \num{2.835e-3} & 0.886 \\
\bottomrule
\end{tabular}
\end{table*}

\begin{table*}[htbp]
\centering
\caption{Temperature dependence of the direct, Raman, and total relaxation rates (arbitrary units), and the corresponding local power-law exponent $\eta(T)$.}
\label{tab:T1_rates_cuoep}
\begin{tabular}{ccccc}
\toprule
$T$ (K) & $(1/T_1)_\mathrm{direct}$ & $(1/T_1)_\mathrm{Raman}$ & $(1/T_1)_\mathrm{total}$ & $\eta(T)$ \\
\midrule
10  & \num{1.8636e-12} & \num{2.4883e-14} & \num{1.8884e-12} & --    \\
15  & \num{1.3886e-11} & \num{2.5132e-12} & \num{1.6400e-11} &  6.51 \\
20  & \num{1.2554e-10} & \num{4.6376e-11} & \num{1.7192e-10} & 10.39 \\
30  & \num{4.6119e-9}  & \num{1.8253e-9}  & \num{6.4372e-9}  &  6.50 \\
50  & \num{1.1365e-7}  & \num{1.0107e-7}  & \num{2.1472e-7}  &  5.67 \\
80  & \num{8.3274e-7}  & \num{2.0886e-6}  & \num{2.9213e-6}  &  5.71 \\
100 & \num{1.7572e-6}  & \num{7.3014e-6}  & \num{9.0586e-6}  &  4.90 \\
150 & \num{5.5394e-6}  & \num{5.6587e-5}  & \num{6.2126e-5}  &  4.45 \\
200 & \num{1.1230e-5}  & \num{2.1289e-4}  & \num{2.2412e-4}  &  4.36 \\
300 & \num{2.7905e-5}  & \num{1.2344e-3}  & \num{1.2623e-3}  & --    \\
\bottomrule
\end{tabular}
\end{table*}

\begin{table*}[htbp]
\centering
\caption{LASSO fit-quality summary for all five $g$-tensor channels. $R^2_\mathrm{shuf}$ denotes the coefficient of determination under shuffled cross-validation. $n_\alpha$ and $n_\beta$ report the number of nonzero linear and bilinear coefficients out of the total available ($n_\mathrm{low}=33$ linear features; 561 bilinear features).}
\label{tab:lasso_fit_quality_cutpp}
\begin{tabular}{lcccccc}
\toprule
Channel & $\lambda_\mathrm{best}$ & $R^2_\mathrm{shuf}$ & $n_\alpha$ (/33) & $n_\beta$ (/561) & $\sigma_\mathrm{signal}$ & Expl.\ frac.\ \\
\midrule
$xx$        & \num{3.915e-7} & 0.9774 & 11 & 223 & \num{1.076e-3} & 0.850 \\
$yy$        & \num{4.450e-7} & 0.9793 & 10 & 213 & \num{1.139e-3} & 0.856 \\
$zz$        & \num{1.421e-6} & 0.9809 & 13 & 227 & \num{3.754e-3} & 0.862 \\
$\perp$     & \num{2.446e-7} & 0.9911 & 10 & 238 & \num{9.620e-4} & 0.906 \\
$\mathrm{iso}$ & \num{9.479e-7} & 0.9798 & 10 & 221 & \num{1.867e-3} & 0.858 \\
\bottomrule
\end{tabular}
\end{table*}

\begin{table*}[htbp]
\centering
\caption{Temperature dependence of the direct, Raman, and total relaxation rates (arbitrary units), and the corresponding local power-law exponent $\eta(T)$.}
\label{tab:T1_rates_cutpp}
\begin{tabular}{ccccc}
\toprule
$T$ (K) & $(1/T_1)_\mathrm{direct}$ & $(1/T_1)_\mathrm{Raman}$ & $(1/T_1)_\mathrm{total}$ & $\eta(T)$ \\
\midrule
10  & \num{4.2034e-9} & \num{8.8118e-11} & \num{4.2915e-9} & --   \\
15  & \num{2.2582e-8} & \num{1.5642e-9}  & \num{2.4146e-8} & 3.92 \\
20  & \num{5.6352e-8} & \num{8.4725e-9}  & \num{6.4825e-8} & 3.57 \\
30  & \num{1.6556e-7} & \num{6.9142e-8}  & \num{2.3470e-7} & 3.11 \\
50  & \num{5.7338e-7} & \num{8.7384e-7}  & \num{1.4472e-6} & 4.03 \\
80  & \num{1.8858e-6} & \num{1.0775e-5}  & \num{1.2660e-5} & 5.48 \\
100 & \num{3.3263e-6} & \num{3.5114e-5}  & \num{3.8440e-5} & 4.91 \\
150 & \num{8.8856e-6} & \num{2.6057e-4}  & \num{2.6946e-4} & 4.51 \\
200 & \num{1.7063e-5} & \num{9.6951e-4}  & \num{9.8657e-4} & 4.40 \\
300 & \num{4.0845e-5} & \num{5.5774e-3}  & \num{5.6183e-3} & --   \\
\bottomrule
\end{tabular}
\end{table*}

\begin{table*}[htbp]
\centering
\caption{LASSO fit-quality summary for all five $g$-tensor channels. $R^2_\mathrm{shuf}$ denotes the coefficient of determination under shuffled cross-validation. $n_\alpha$ and $n_\beta$ report the number of nonzero linear and bilinear coefficients out of the total available ($n_\mathrm{low}=38$ linear features; 741 bilinear features).}

\label{tab:lasso_fit_quality}
\begin{tabular}{lcccccc}
\toprule
Channel & $\lambda_\mathrm{best}$ & $R^2_\mathrm{shuf}$ & $n_\alpha$ (/38) & $n_\beta$ (/741) & $\sigma_\mathrm{signal}$ & Expl.\ frac.\ \\
\midrule
$xx$        & \num{8.328e-7} & 0.9868 & 13 & 215 & \num{1.814e-3} & 0.885 \\
$yy$        & \num{8.149e-7} & 0.9897 & 13 & 222 & \num{1.971e-3} & 0.898 \\
$zz$        & \num{2.685e-6} & 0.9878 & 14 & 209 & \num{6.216e-3} & 0.890 \\
$\perp$     & \num{8.172e-7} & 0.9921 & 13 & 199 & \num{1.835e-3} & 0.911 \\
$\mathrm{iso}$ & \num{1.430e-6} & 0.9895 & 15 & 215 & \num{3.233e-3} & 0.897 \\
\bottomrule
\end{tabular}
\end{table*}

\begin{table*}[htbp]
\centering
\caption{Temperature dependence of the direct, Raman, and total relaxation rates (arbitrary units), and the corresponding local power-law exponent $\eta(T)$.}
\label{tab:T1_rates}
\begin{tabular}{ccccc}
\toprule
$T$ (K) & $(1/T_1)_\mathrm{direct}$ & $(1/T_1)_\mathrm{Raman}$ & $(1/T_1)_\mathrm{total}$ & $\eta(T)$ \\
\midrule
10  & \num{1.6944e-9} & \num{1.3539e-16} & \num{1.6944e-9} & --   \\
15  & \num{8.3901e-9} & \num{9.7782e-14} & \num{8.3902e-9} & 3.69 \\
20  & \num{2.1829e-8} & \num{4.6922e-12} & \num{2.1834e-8} & 3.66 \\
30  & \num{8.6222e-8} & \num{6.5179e-10} & \num{8.6873e-8} & 3.32 \\
50  & \num{4.8896e-7} & \num{8.1963e-8}  & \num{5.7092e-7} & 3.84 \\
80  & \num{2.0333e-6} & \num{2.2641e-6}  & \num{4.2974e-6} & 5.13 \\
100 & \num{3.7976e-6} & \num{8.5168e-6}  & \num{1.2314e-5} & 4.72 \\
150 & \num{1.0866e-5} & \num{7.1429e-5}  & \num{8.2295e-5} & 4.46 \\
200 & \num{2.1533e-5} & \num{2.7685e-4}  & \num{2.9839e-4} & 4.38 \\
300 & \num{5.2941e-5} & \num{1.6410e-3}  & \num{1.6939e-3} & --   \\
\bottomrule
\end{tabular}
\end{table*}

\clearpage

\bibliography{refs}

\end{document}